\documentclass[useAMS,usenatbib,usedcolumn,usegraphicx]{mn2e}
\usepackage{times}
\usepackage{longtable}
\usepackage{tabularx}

\title{X-ray source populations in the Galactic Plane}

\author[A.\,D.\,P. Hands et al.]
{A.\,D.\,P. Hands$^{1}$, R.\,S. Warwick$^{1}$, M.\,G. Watson$^{1}$ 
\& D.\,J. Helfand$^{2}$\\
$^{1}$Department of Physics and Astronomy, University 
of Leicester, University Road, Leicester, LE1 7RH \\
$^{2}$Columbia Astrophysics Laboratory, Columbia University, 
550 West 120th Street, New York, NY 10027, USA}
\date{}

\def\xmm{{\it XMM-NEWTON~\/}}

\def\asca{{\it ASCA~\/}}

\def\chan{{\it CHANDRA~\/}}
\def\sax{{\it BeppoSAX~\/}}
\def\xmmc{{\it XMM-NEWTON,~\/}}

\def\ergseccm{{$\rm ~erg~s^{-1}~cm^{-2}$}}
\def\ergseccmdeg{{$\rm ~erg~s^{-1}~cm^{-2}~deg^{-2}$}}
\def\ergsec{{$\rm ~erg~s^{-1}$}}

\def\ctsecdeg{{$\rm ~count~s^{-1}~deg^{-2}$}}

\def\eg{{\it e.g.,~\/}}

\def\deg{{${ }^{\circ}$}}
\newcommand{\citeb}[1]{\citeauthor{#1}~\citeyear{#1}}

\begin{document}

\maketitle

\begin{abstract}

We present the first results from the \xmm Galactic Plane Survey (XGPS).  
In the first phase of the programme, 22 pointings were used to cover a region 
of approximately three square degrees between 19\deg -- 22\deg in 
Galactic longitude and $\pm$0.6\deg in latitude.  In total we have resolved 
over 400 point X-ray sources, at $\geq 5 \sigma$ significance,   
down to a flux limit of $\sim2 \times 10^{-14}$ \ergseccm (2--10~keV).  
The sources exhibit a very wide range of spectral hardness,  with
interstellar absorption identified as a major influence. The source 
populations detected in the soft (0.4--2 keV) band and hard (2--6 keV) band 
show surprisingly little overlap.  The majority of the soft sources appear 
to be associated with
relatively nearby stars with active stellar coronae, judging from their
high coincidence with bright stellar counterparts. 

The combination of the XGPS measurements in the hard X-ray band with the 
results from  earlier surveys carried out by \asca and 
\chan reveals the form of the low-latitude X-ray source counts
over 4 decades of flux. It appears that extragalactic sources dominate below
$\sim10^{-13}$ \ergseccm (2--10~keV), with a predominantly
Galactic source population present above this flux threshold.
The nature of the faint Galactic population observed by \xmm remains uncertain,
although cataclysmic variables and RS CVn systems may contribute substantially.
\xmm observes an enhanced surface brightness in the Galactic plane
in the 2--6 keV band associated with Galactic Ridge X-ray Emission (GRXE). 
The integrated contribution of Galactic sources plus the breakthrough of 
extragalactic signal accounts for up to 20 per cent of the observed 
surface brightness.  The XGPS results are consistent with the picture
suggested from a deep \chan observation in the Galactic plane, namely that 
the bulk of the GRXE is truly diffuse.

\end{abstract}


\section{Introduction}

With the current generation of X-ray astronomy missions, we are  for the first 
time able to carry out high sensitivity, coherent surveys of selected regions 
of the Galactic plane.  In particular, the \xmm mirrors afford a large 
collecting area ($\sim 4650 \rm~cm^{-2}$ total geometric area) with 
good spatial resolution (FWHM $\sim6''$ and HEW $\sim15''$ on-axis) 
over a wide field of view ($30'$ diameter). In 
combination with the EPIC CCD cameras, this provides an excellent 
facility for surveying sky regions subtending many square degrees
down to relatively faint flux levels in both the hard ($> 2$ keV) and soft 
($< 2$ keV) X-ray bands.

The goal of the \xmm Galactic Plane Survey (XGPS) is two-fold. The first 
objective is to study the properties of the Galactic X-ray source population 
at intermediate flux levels (down to $\sim 2 \times 10^{-14}$ \ergseccm~in the 
2--10~keV band  but an order of magnitude fainter in flux terms in the softer 
0.4--2~keV band). The second is to search for extended, low
X-ray surface brightness features including variations in the underlying 
diffuse Galactic Ridge X-ray Emission (GRXE; \citeb{a255};
\citeb{n317}; \citeb{pasj38}; \citeb{a404}; \citeb{a491}; \citeb{a505}). 

The nature of the X-ray source population at high X-ray fluxes was
established by early all-sky surveys and subsequent identification
programmes, which revealed that the brightest sources in our Galaxy are
predominantly X-ray binaries and supernova remnants.
More sensitive surveys of the Galactic plane have since been made, including
those made by {\it ROSAT} (\citeb{aa246}) and 
{\it ASCA} (\citeb{a134}) complemented by 
the serendipitous surveys carried out with the Einstein observatory 
(\citeb{a278}). Together, these surveys have provided some glimpses of the
X-ray source population at lower X-ray fluxes, and hence effectively at
lower X-ray luminosities for Galactic objects, although the picture is far
from complete. At soft X-ray energies ($<2$ keV) {\it ROSAT} studies in 
particular
have shown that coronal emission from relatively nearby active stars
dominates (\eg {\citeb{aa246b}). Above 2 keV the characteristics of the harder 
population are far 
less well-defined, although it is clear that accreting binary sources 
(both X-ray binaries and cataclysmic variables) make a significant 
contribution.

To date the XGPS  survey has been
targeted  at several locations in the Galactic segment between the Galactic 
Centre and the Scutum Spiral Arm. Here we report the results from the first 
phase of the XGPS (hereafter XGPS-I), which has entailed a total of 22 
\xmm pointings, covering a region of approximately three square degrees 
between 19\deg -- 22\deg
in Galactic longitude and $\pm$0.6\deg in latitude. Over 400 discrete 
point-like X-ray sources have been detected in XGPS-I and in this paper we 
focus on the properties of this source population and the contribution these 
discrete sources make to the GRXE. In a second paper (Hands et al. 2004, in 
preparation) we will present the results of a search for low-surface 
brightness, spatially extended X-ray sources in the XGPS-I fields and also 
report on the properties of the underlying diffuse GRXE.

\begin{table*}
\begin{center}
\caption{Observation log for the XGPS-I. \label{tab:log}}

\vspace{0.1in}
\begin{tabular}{lcllrrrr} \hline

Field & & \multicolumn{2}{c}{Field Centre (J2000)} & MOS & pn & MOS & pn\\
Name  & Observation Date & R.A.   & Dec. 
& exposure$^a$  & exposure$^b$ & fraction$^c$ & fraction$^c$ \\ \hline

Ridge 1 & 2000-10-08 & 18 26 00.4 & -12 14 55.9 & 8393  & 5914 & 0.95 & 0.85 \\
Ridge 2 & 2002-09-21 & 18 26 48.4 & -11 52 48.7 & 13667 & 12046 & 1.00 & 1.00 \\
Ridge 3 & 2000-10-11 & 18 27 36.4 & -11 30 40.4 & 12044 & 9648 & 0.95 & 0.85 \\
Ridge 4 & 2000-10-12 & 18 28 17.0 & -11 07 53.0 & 9256  & 12998 & 0.94 & 0.42 \\
Ridge 5 & 2002-09-17 & 18 29 06.0 & -10 45 03.0 & 13667 & 12046 & 0.98 & 0.93 \\
XGPS 1  & 2001-03-08 & 18 25 04.6 & -11 50 00.4 & 7794  & 4797 & 1.00 & 1.00 \\
XGPS 2  & 2001-03-10 & 18 27 34.0 & -12 09 20.0 & 9144  & - & 1.00 & - \\
XGPS 3  & 2001-03-10 & 18 25 49.0 & -11 28 42.7 & 9144  &  - & 1.00 & - \\
XGPS 4  & 2001-03-10 & 18 28 19.4 & -11 48 05.2 & 9144  &  - & 1.00 & - \\
XGPS 5  & 2001-03-22 & 18 26 35.7 & -11 07 32.1 & 9994  & 7348 & 0.96 & 0.95 \\
XGPS 6  & 2001-03-22 & 18 29 05.9 & -11 26 57.4 & 8794  & 6148 & 0.44 & 0.23 \\
XGPS 7  & 2001-03-24 & 18 27 21.2 & -10 46 17.8 & 7637  & 4998 & 0.95 & 0.98 \\
XGPS 8  & 2002-09-29 & 18 29 50.8 & -11 05 41.2 & 13167 & 11546 & 1.00 & 1.00 \\
XGPS 9  & 2001-03-26 & 18 28 06.4 & -10 25 10.0 & 11894 & 9248 & 0.92 & 0.62 \\
XGPS 10 & 2001-10-03 & 18 30 36.2 & -10 44 29.3 & 9767  & 8146 & 0.97 & 0.83 \\
XGPS 11 & 2002-09-19 & 18 29 43.9 & -10 24 09.6 & 8409  & 6788 & 0.92 & 0.81 \\
XGPS 12 & 2002-09-27 & 18 28 59.1 & -10 06 50.9 & 7367  & 5746 & 1.00 & 1.00 \\
XGPS 13 & 2003-04-10 & 18 31 25.5 & -10 24 32.5 & 6666  & 5047 & 0.73 & 0.66 \\
XGPS 14 & 2002-03-11 & 18 30 29.3 & -10 02 47.1 & 7269  & 4998 & 1.00 & 1.00 \\
XGPS 15 & 2002-03-27 & 18 29 36.6 & -~9 42 41.1 & 7274  & 4998 & 0.82 & 0.96 \\
XGPS 16 & 2002-03-15 & 18 32 06.5 & -10 01 51.8 & 7762  & 5486 & 1.00 & 1.00 \\
XGPS 17 & 2002-03-29 & 18 31 13.9 & -~9 41 39.4 & 7274  & 4998 & 0.99 & 0.94 \\ 
\hline
        &            &           &           &       & \\
\multicolumn{4}{l}{$^a$ Total exposure for the MOS 1 camera (s)} \\
\multicolumn{4}{l}{$^b$ Total exposure for the pn camera (s)} \\
\multicolumn{4}{l}{$^c$ Fraction of the exposure time used in producing
images} \\

\end{tabular}
\end{center}

\end{table*}

\section{Observations}

The XGPS-I programme comprises 22 \xmm pointings carried out during 
the period between October 2000 and April 2003 (see Table \ref{tab:log}).
Five of these observations formed part of the SSC Guaranteed Time programme 
(the Ridge 1-5 fields), whereas the remaining time was awarded to
an AO1 programme (PI: Warwick; the XGPS 1-17 fields). The allocated exposure 
times for these two sets of observation were 9 ks and 5 ks respectively,
although in most instances somewhat longer exposure times were actually 
scheduled (see Table \ref{tab:log}). In all cases the EPIC cameras were
operated in {\it Full Window Mode}  with the medium filter selected. 
In the event, the completion of this survey proved problematic due to the 
impact of intervals of high instrumental background\footnote{The background
enhancements are attributed to a 
highly variable flux of soft protons which in orbit appear to be
channeled by the X-ray mirrors onto the CCD detectors.}
on the data quality, at least for a subset of the pointings. Several of the 
pointings were in fact repeated so to mitigate the worst effects of
this contamination (Table \ref{tab:log} refers to the observations 
actually used in the present work).

\section{Data Analysis}
\label{sec:data}

\subsection{Data screening and image extraction}

We have analysed the X-ray data from the three EPIC cameras on \xmmc two of
which incorporate MOS CCDs (MOS 1 and MOS 2) (\citeb{aa365}) and one based
on pn technology (\citeb{aa365p}).  As noted earlier the instrument 
background in both the MOS and pn cameras is highly variable with the 
transition from 
quiescent conditions to a severe flaring episode often occurring on timescales
shorter than $\sim1000$ s and involving an increase in the background
count rate by factors ranging from a few up to several orders of 
magnitude. 
In order to assess the background conditions in each of the XGPS-I
observations we have extracted the full-field light curve for
events with energy in the range 0.2--12 keV. The results are illustrated
in Fig. \ref{fig:lcurves} for the MOS 1 camera.

\begin{figure*}
\begin{center}
\includegraphics[height=6in,width=9in,angle=270]{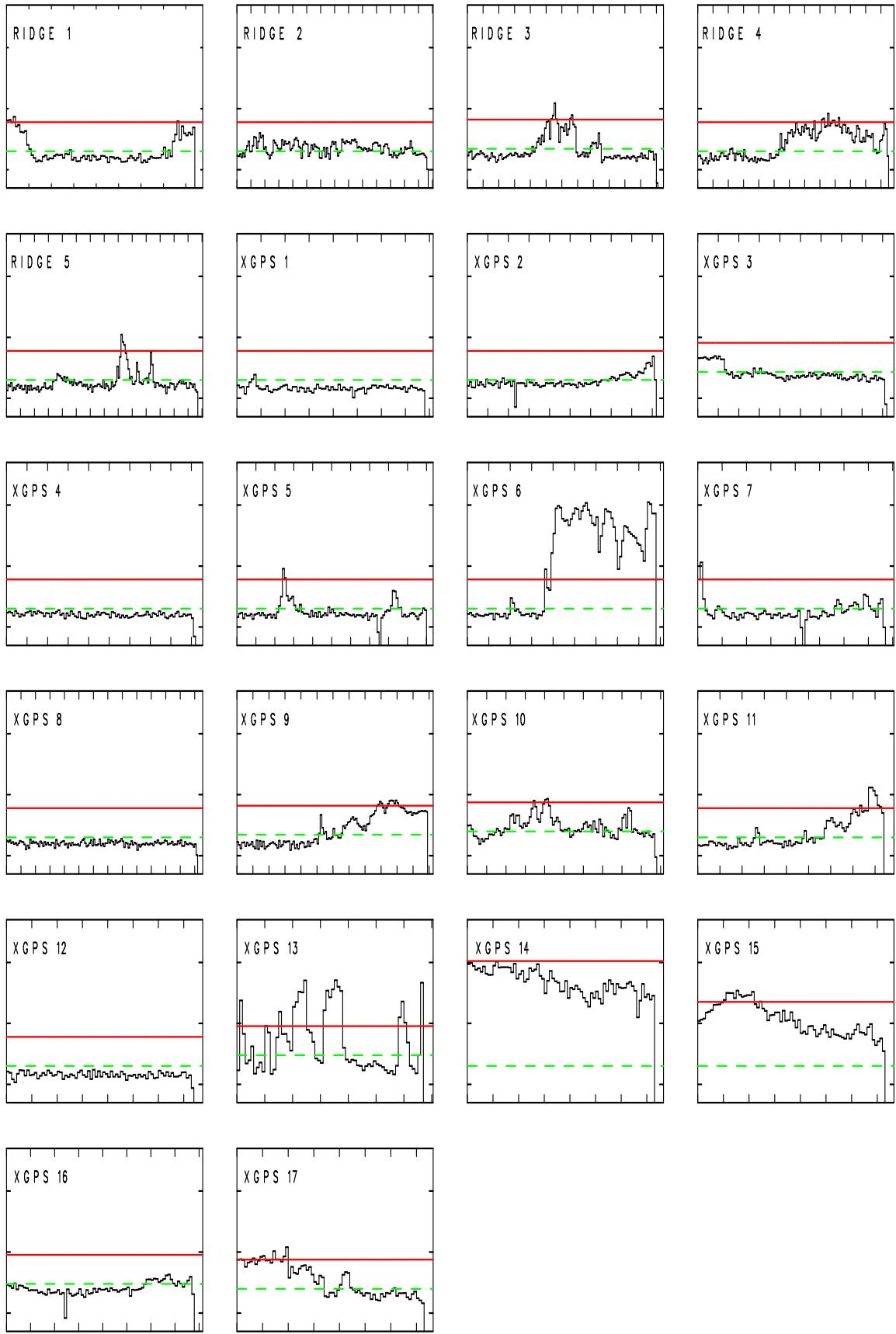}
\caption{MOS 1 full-field light curves for all 22 XGPS-I observations. 
In each case the units of the vertical axis are counts per 100 s in the 
full 0.2--12 keV band.  The scaling is logarithmic with the three tick 
marks corresponding to 100, 1000 and 10000 counts per 100 s, respectively.
The unit of the horizontal axis is time with tick marks every 1000 s. 
The thresholds used to 
exclude the most intense background flares are shown in each case 
(solid line). The more stringent thresholds used to extract
datasets with ``near-quiescent'' background conditions are also 
indicated (dashed line). 
\label{fig:lcurves}}
\end{center}
\end{figure*}

Prior to constructing X-ray images, the X-ray events\footnote{We take as a 
starting point
in our analysis the calibrated event list produced by the standard \xmm
pipeline, together with exposure maps and other pipeline products. We
select X-ray events corresponding to patterns 0--12 for the MOS and 
0--4 in the pn.} recorded in each EPIC camera for each observation
must be filtered in various ways.  
The first step was to use the full-field light curve to exclude time 
intervals when the instrument background was unduly high.  The fact that 
the XGPS-I observations are relatively short ({\it i.e.,} 6-14 ks actual 
exposure time) and the background flaring episodes are rather common and 
have durations of several ks or longer, means that ``near-quiescent'' 
background conditions are typically experienced only for a fraction of the 
total on-time 
(with this fraction approaching zero in some observations). As a compromise
between selecting clean data on the one hand and having sufficient source 
counts to make source detection effective on the other, we set a threshold for
data exclusion at a full-field count rate roughly three times higher than 
the ''lowest level'' of the observation (see Fig. \ref{fig:lcurves} and
Table \ref{tab:log}). For
most of the observations this resulted in a cut at a rather similar 
count-rate setting. However, the XGPS 14 and XGPS 15 observations were 
subject to enhanced background levels throughout the exposure and for these 
a significantly higher threshold count-rate was required (implying
a much reduced sensitivity to faint cosmic X-ray sources in these 
observations).
For some purposes (eg. searching for low-surface brightness
X-ray features within a particular observation or looking for
variations in the underlying GRXE across many fields) a much more 
stringent rejection of high background intervals is required.
The thresholds used to identify near-quiescent background 
conditions are also shown in Fig. \ref{fig:lcurves}. All the data
from the XGPS 14 and XGPS 15 observations were rejected for this analysis.

This empirical approach is broadly similar to that later adopted for the
construction of the first \xmm catalogue (\citeb{an324x}).

Once the temporal filter has been applied, the next step is to make images 
in specific energy bands for further analysis.  Here we have used
a soft (0.4--2.0 keV) band, a hard (2.0--6.0 keV) band and a broad band 
representing the combination of the soft and hard channels (0.4--6 keV). Our
choice of bands was made to optimise detection signal-to-noise. Note that it
differs somewhat from that used in the standard \xmm data 
products.
In 
the case of the MOS cameras we specifically excluded two narrow energy bands 
which are contaminated by fluorescent Al and Si lines originating within the
detector\footnote{The energy 
ranges excluded  were 1.4--1.575 keV and 1.675--1.8 keV 
corresponding to the K$_\alpha$ lines of neutral Al and Si respectively.}
(see Fig. \ref{fig:mosedges}).  
Although the Si line is not a prominent feature in the pn background, for 
consistency we use the same energy band selection
for the pn data.

\begin{figure}
\begin{center}
\includegraphics[height=2.5in,width=!,angle=270]{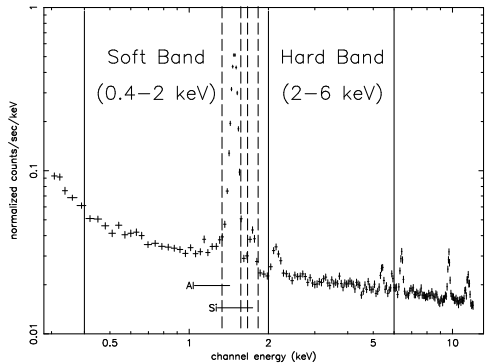}

\caption{The spectra collected from edge regions of the MOS CCDs 
which lie underneath the filter wheel and hence are not exposed to the
X-ray sky.  The solid lines define the boundaries of the 
energy bands used in the analysis. The two regions identified by the 
dashed lines, which encompass the fluorescence lines of Al and Si produced 
within the MOS cameras, were specifically excluded from the soft band.
Although other instrument fluorescence lines are present in the hard
band, they are of relatively low intensity and have very little impact
on the hard-band background count rates. 
\label{fig:mosedges}}

\end{center}
\end{figure}

The final filter used to select the data is a spatial discriminator to 
distinguish between the areas of the CCDs which are exposed to the sky and 
those which are not. This is particularly important for the two MOS cameras 
for which a significant fraction of the area of the outer CCDs is shielded by 
the camera's filter wheel.  The spatial mask used to perform this 
filtering was derived from a central band exposure map (2.0--4.5 keV), 
produced by the standard SAS pipeline procedure. In practice,  we 
also applied additional masking so as to exclude the regions of the
field of view where, due to the mirror vignetting, the effective exposure 
was less than 25 per cent of the on-axis value. 

At this stage we were utilising 3 energy bands (soft, hard \&
broad) per EPIC camera (MOS 1, MOS 2, pn), to give a total of nine 
separate X-ray images per observation, with each image consisting of a 
$600 \times 600$ array of $4''$ pixels. In order to improve the 
signal-to-noise 
ratio we subsequently co-added each pair of MOS images. In carrying
out source detection we have treated the MOS and pn data as completely 
separate channels, which can be compared for quality control purposes. 

Fig. \ref{fig:expmaps} illustrates the sky coverage of the survey
for both the MOS and pn cameras in the form of a mosaic of
the exposure maps from the individual XGPS-I pointings. In effect the 
survey uses three rows of pointings in a close-packed hexagonal pattern (with 
a spacing between adjacent field centres of $24'$), so as to give efficient
(but not particularly uniform) coverage of a narrow strip of the Galactic 
plane. 

\begin{figure*}
\begin{center}
\includegraphics[height=6in,width=!,angle=270]{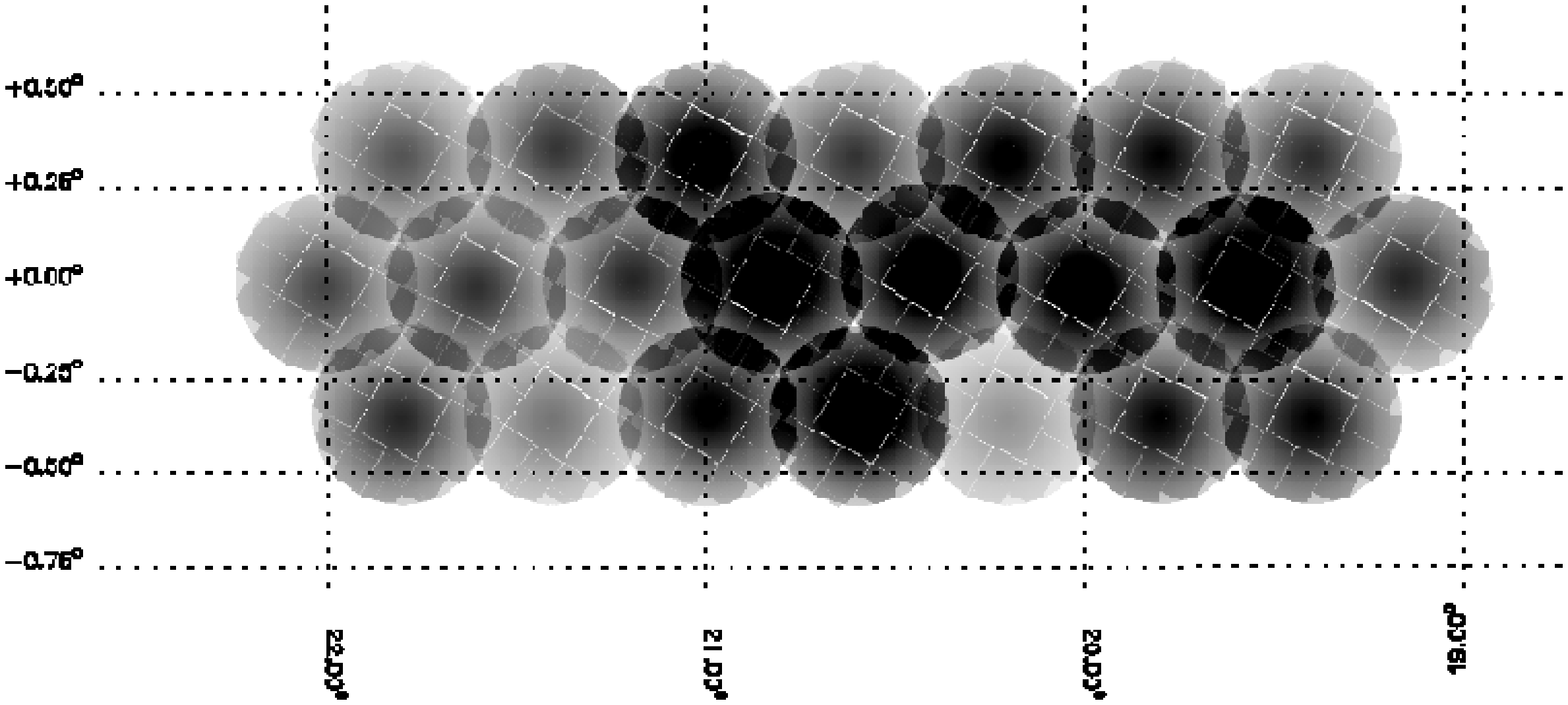}
\includegraphics[height=6in,width=!,angle=270]{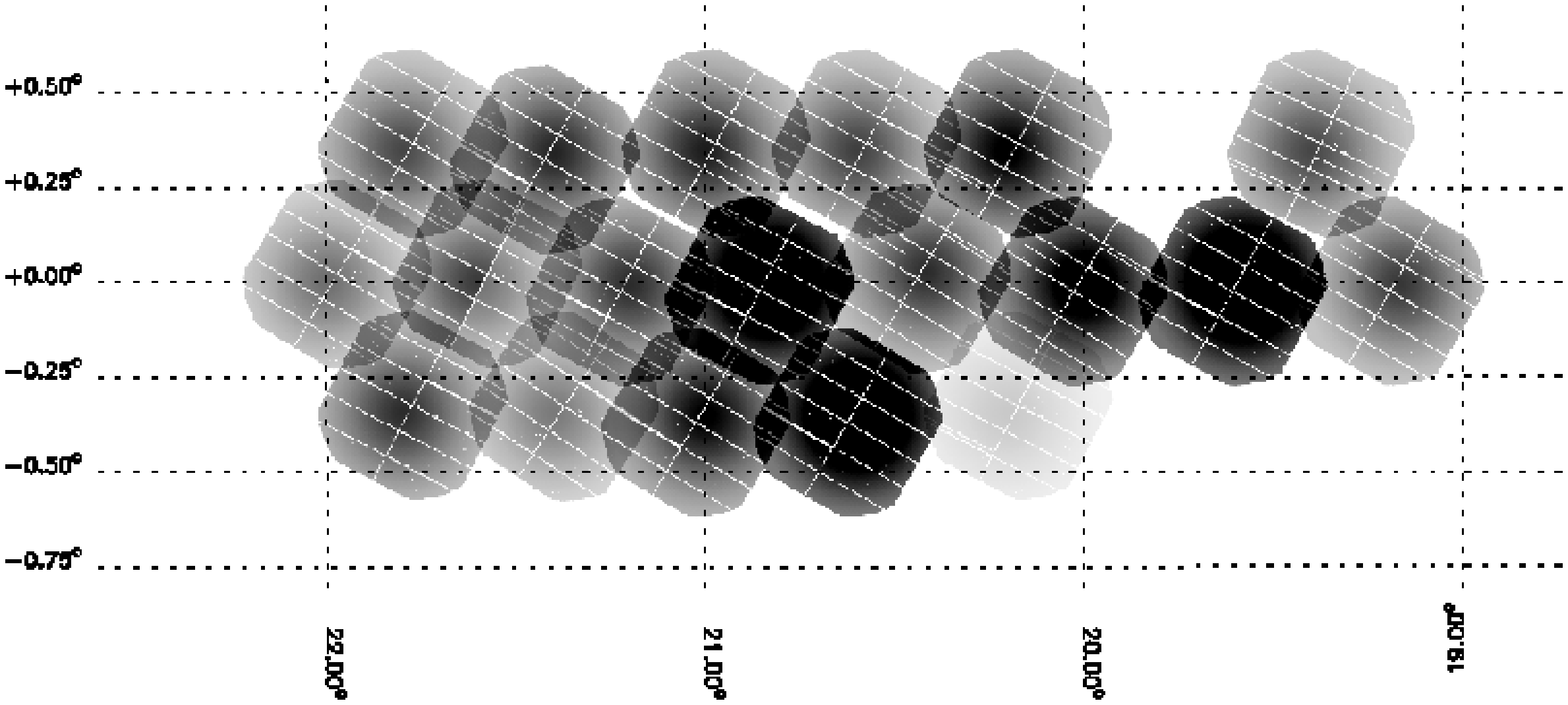}

\caption{{\it Top panel:} A mosaiced (2--4.5 keV) exposure map showing the 
sky coverage provided by the MOS cameras over the full set of XGPS-I
observations. The image is plotted in Galactic coordinates and employs a 
simple rectangular projection. The grey-scale 
(darker for longer exposures) corresponds
to the accumulated (MOS 1 + MOS 2) exposure time at different points in
the survey region, with the variation for each individual pointing
largely reflecting the vignetting function of the \xmm mirrors.
{\it Bottom panel:} The same information for the pn camera. The gaps in the
pn exposure map correspond to XGPS-I observations for which 
pipeline-processed pn data are not available. The maximum exposure
is 18 ks and 6.5 ks in the 
MOS and pn images respectively.
\label{fig:expmaps}}

\end{center}
\end{figure*}

\begin{figure*}
\begin{center}
\includegraphics[height=6in,width=!,angle=270]{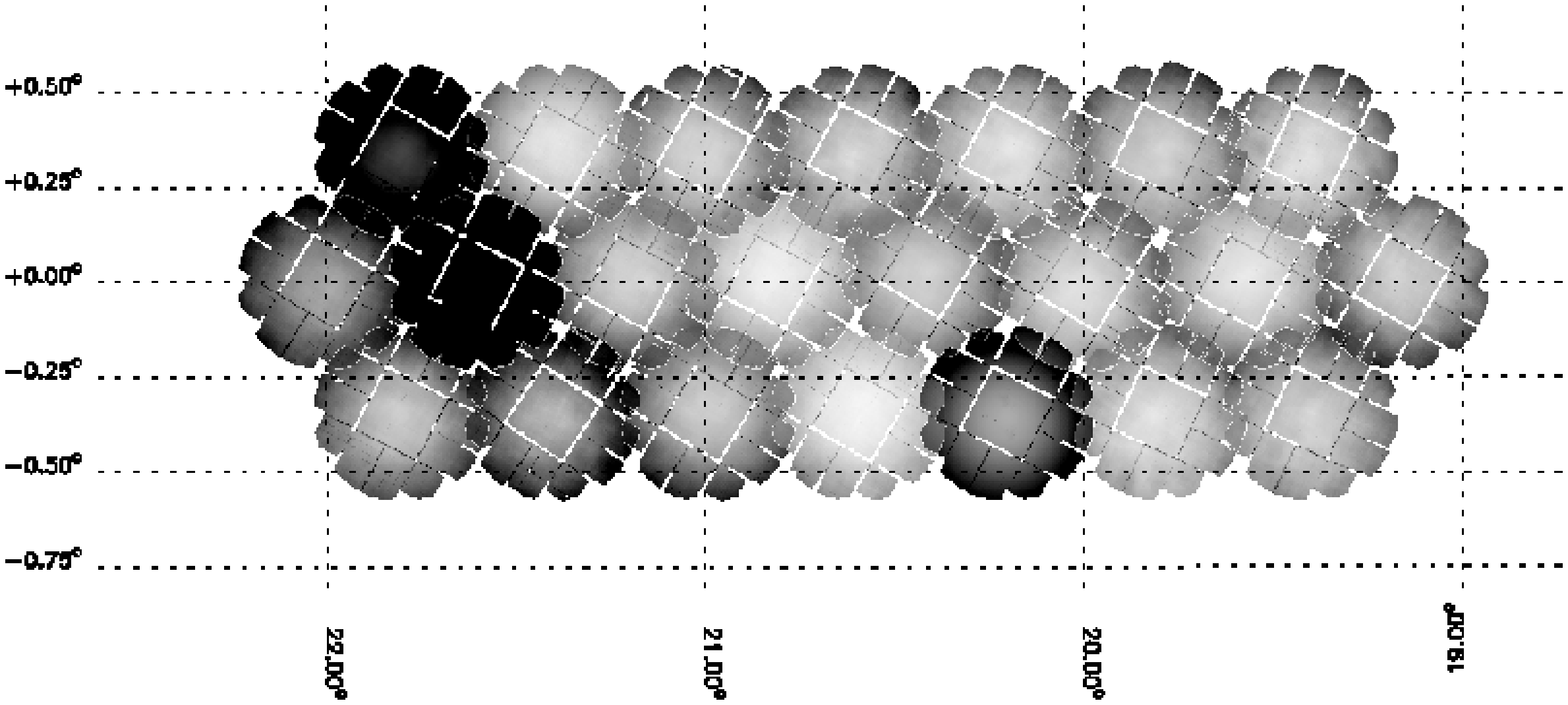}
\includegraphics[height=6in,width=!,angle=270]{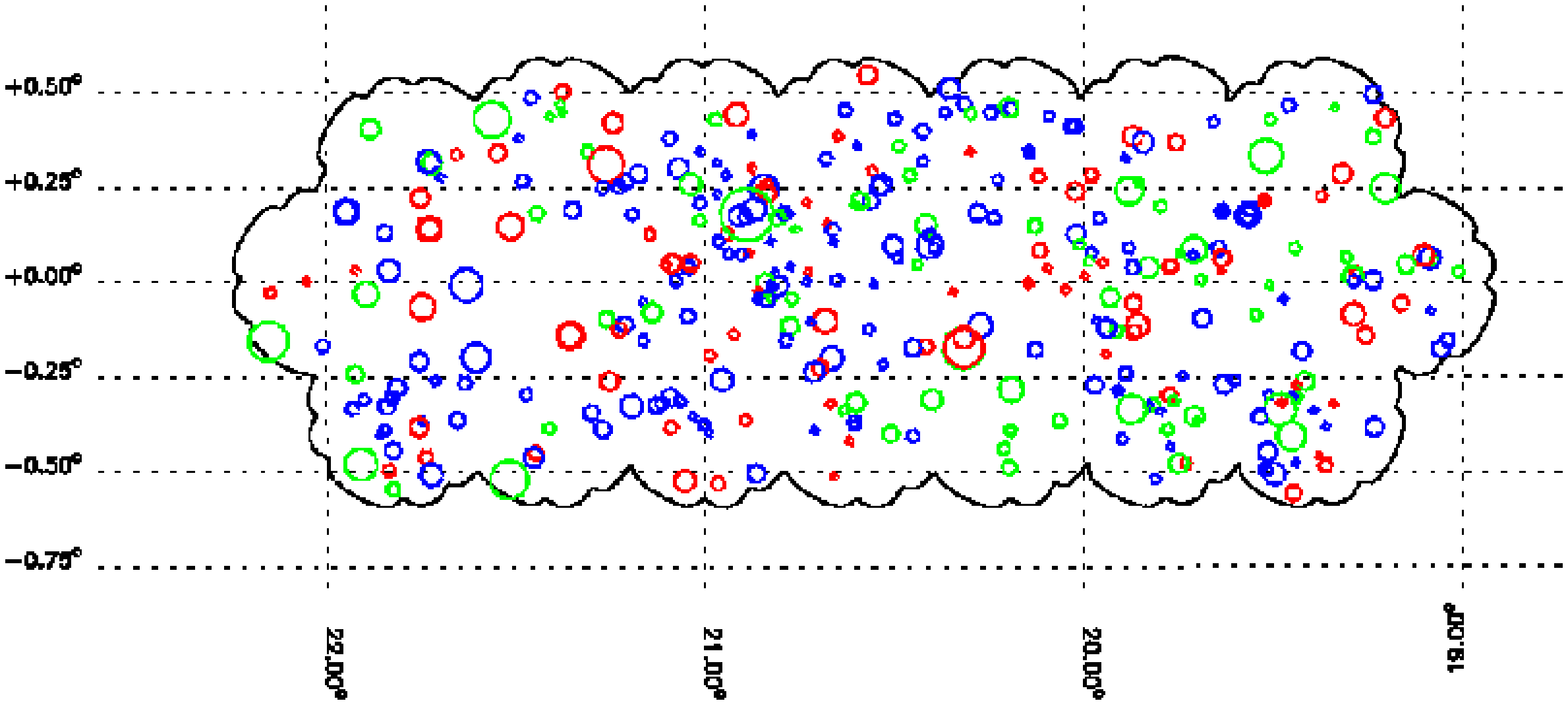}

\caption{{\it Top Panel:} Mosaic showing the variation in source detection
sensitivity across the 22 XGPS-I fields.  The broad band
(0.4--6 keV) source count rate required to give a $5\sigma$ detection varies 
from $0.5~\rm cts~ks^{-1}$ (white) to $3~\rm cts~ks^{-1}$ (black)
across the survey region dependent on the exposure time and particle 
background rate.
{\it Bottom Panel:} Schematic image showing the sources detected 
by the MOS cameras in the soft (0.4--2 keV), hard (2--6 keV) and broad
(0.4--6 keV) energy bands.  Blue circles represent sources
detected in the hard band but not the soft band; red circles represent sources
detected in the soft band but not the hard band; green circles represent the
remaining sources which are either detected in the soft and hard bands or in
the broad band alone.  The radius of each circle is a logarithmic measure of 
the count rate of source.
\label{fig:images}}
\end{center}
\end{figure*}

\subsection{Detecting sources}
\label{sec:detect}

We have employed an iterative sliding-box algorithm to detect point sources 
in the individual XGPS-I fields\footnote{Our approach was developed before
the standard SAS source detection chain was mature enough to be considered
for this project. Although our procedures are simpler than those
incorporated in the SAS chain used in standard \xmm processing, we do not
expect that a detailed comparison would reveal any major differences in
results and in particular our approach 
to background estimation is
better-matched to the specific background issues faced at low Galactic
latitudes and in low exposures.}. Briefly the process was as follows. The raw 
image was lightly smoothed and used to identify bright regions corresponding 
to individual point sources or groups of sources.  These source regions were
then excised and the remaining data heavily smoothed to produce a 
sky-background map\footnote{The light smoothing was performed using a circular
Gaussian mask with $\sigma = 8''$ (1 pixel = $4''$), 
whereas the mask 
used for heavy smoothing was a square top-hat function of dimension $160''$.}.
After subtracting the derived background map from the raw image, the 
data were again lightly smoothed and then scanned for local peaks above a 
specified surface brightness threshold. At the position of each local 
peak, we use a cell of radius $16''$ 
to extract a source plus background counts value (C) from 
the original raw image and a corresponding background estimate (B)
from the background map. The net
count from the source (S) was then given simply as $S = C-B$.  
In order for a peak to qualify as a source detection, two criteria 
were applied, namely S $\geq 10$ and S $\geq 5 \surd B$, implying
a detection significance of $ \geq 5 \sigma$.
The list of sources detected by this method were subsequently used to
define a new source mask and a revised background map. A further iteration 
of the source detection procedure then resulted in a final source list
for the observation in question.
Since the sensitivity to point sources depends on both the
exposure time and the particle background rate, the sensitivity
map of the full survey shows a somewhat different spatial variation 
to that of the corresponding exposure map (Fig. \ref{fig:images}, top panel).

The process described above produces 6 separate source lists per XGPS-I 
observation (two detector channels and three spectral bandpasses).  These
source lists were merged by correlating all sources lying within 
$20''$ of each other.  The position offsets for detections 
of the same source in two detector channels or in different bandpasses 
of the same detector were found to be distributed such that 68\% (90\%)
were contained within a radius of $\approx 2.8''$ ( $\approx 4''$),
which is consistent with the estimated statistical
errors on the positions for sources at the faint end of the brightness
distribution. However, a much broader correlation region was chosen so 
that sources would not be identified as 
distinct if the separation between them was less than or comparable to the
half energy width of the EPIC point spread function. Fig. \ref{fig:mospn} 
shows a comparison of the MOS and pn broad-band detections in a central part 
of the Ridge 3 observation.  In total four sources are detected in this
sub-region but interestingly  only one of them is classed as a detection
in {\it both} the MOS and pn cameras (see \S \ref{catalogue}). 

With the merging process for each observation completed, the next step was 
to select the best
position information for each individual source and use this position
to determine its flux.  Here we used a {\it quality} parameter from the 
source detection process, representing the fraction of ``good 
pixels'' in the source cell.  This parameter is particularly useful
in flagging sources near CCD chip gaps or distorted by bad columns in the 
CCD.  The source position was taken from the (detector/spectral) channel 
with the highest {\it quality}  value. Where more than  one channel had 
{\it quality} $=1$ we arbitrarily used the priority sequence  
pn/broad, MOS/broad, pn/hard, MOS/hard, pn/soft, MOS/soft. 
At the assigned ``best position'' we then extracted counts estimates 
(S=C-B) using both a $16''$ and $24''$ radius cell.  The measured counts were 
subsequently corrected for the signal loss outside the source
cell\footnote{A source cell of radius $16''$ (4 pixels) encompasses 
$\sim 70 \%$ of the counts from a point source (averaged over the field
of view), whereas the corresponding value for a $24''$ (6 pixel) radius cell 
is 80\%. Unless otherwise stated, all the X-ray count rates quoted here 
are corrected for the
signal loss due to the limited size of the source cell.} 
and converted to on-axis count rates using appropriate exposure map 
information. We use the counts
derived from the smaller cells in the source counts analysis, since these are 
matched to the source detection process and the derived sensitivity curves 
(see \S 4.3). For all other purposes we use the counts based on the 
$24''$ radius cells.

The final step was to combine the source lists from the individual
XGPS-I observations into one source catalogue. This involved
the removal of a small number of duplicate entries where sources had
been detected in more than one observation in regions where there was
overlapping coverage; in practice preference was always given to the 
highest sensitivity detection. 

\subsection{Extracting source spectra}
\label{sec:data:spec}

An algorithm was also developed to extract the net spectra of a defined
set of sources in any particular field. In this case a more stringent
version of the background rejection filtering  was employed with the
result that the two XGPS-I observations badly affected
by background flaring were excluded (see Fig. \ref{fig:lcurves}).
The source plus background events were accumulated from within circular 
regions of radius $24''$ centred on the identified source 
positions, with a similarly positioned array of annuli of radius $24-96''$ 
used to extract corresponding background events.
Where two sources occur close to each other, the region of the annulus 
contaminated by the other source was excised.  Background subtraction 
was carried out on a source by source basis and the resulting net-spectra 
summed over the defined set of sources prior to division by the effective 
exposure time. In effect this procedure provides the count-weighted
average spectrum of the set of sources under consideration.

\begin{figure}
\begin{center}
\includegraphics[height=!,width=3in,angle=0]{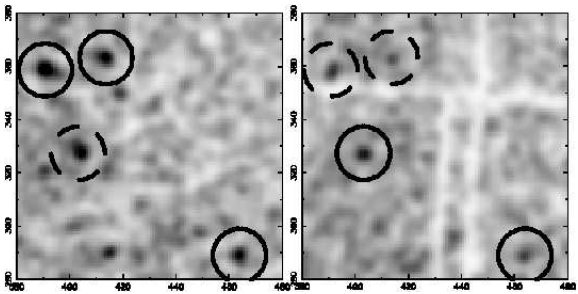}

\caption{Lightly smoothed MOS (left) and pn (right) images of a part 
of the Ridge 3 field  in the broad  (0.4--6 keV) band.  The circles drawn
with solid lines
indicate local peaks (above a specified threshold) which meet the two
source-detection criteria, whereas the circles drawn with dashed lines indicate
peaks which failed this test. A total of four sources are detected
in this subfield, but only 1 source is classed as a significant detection
in {\it both} the MOS and pn cameras. 
\label{fig:mospn}}

\end{center}
\end{figure}

\section{Results}

One of the main aims of the present survey is to study the X-ray source 
population of the Galaxy at relatively faint fluxes.  In this section we 
consider the X-ray 
source catalogue derived from the XGPS-I programme, the X-ray spectral 
properties of the sources and the source count statistics. We also give
brief details of possible optical counterparts based on available
wide-field optical data and other published catalogue information.

\subsection{The XGPS-I Source Catalogue}
\label{catalogue}

A total of 424 discrete X-ray sources satisfied the detection criteria. 
Fig. \ref{fig:images} (bottom panel) illustrates the spatial distribution 
of the XGPS-I sources across the survey region. The full source 
catalogue, including the source positions and the measured count rates
in both the MOS and pn cameras,  is presented in Appendix A.  The X-ray 
spectral hardness ratio (HR)  quoted for each source is defined as:

\begin{eqnarray}
HR & = & \frac{H-S}{H+S} 
\end{eqnarray}

where $H$ is the number of counts measured in the hard band and 
$S$ the corresponding number of soft counts. For sources detected in 
both camera systems we summed the MOS and pn counts in the two 
energy bands before calculating HR. A correction for the differential
vignetting was not applied to the HR value since this was generally
small compared to the statistical error.

The numbers of sources detected in each camera system and in each energy
band are given in Table \ref{tab:detect}. 
Of the 424 sources in the catalogue, 132 are detected in both
the pn and MOS cameras, which represents 59\% of the pn sample
but only 38\% of the MOS sources. The fact that there are 
more  source 
detections in the MOS channel than in the pn camera reflects both the
lack of pn data for some of the fields and also the longer exposures times 
typically achieved for the MOS detectors (the set-up time for the
pn camera is a significant overhead for these rather short observations).
For both the MOS and pn instruments, considerably more sources were detected
in the hard energy band than in the soft band. It is surprising that the 
overlap between the spectral channels, {\it i.e.,} the number of sources 
independently detected in both the hard and soft spectral bands, is so 
small (only $\sim 10-15$\% of the sample). This spectral characteristic
presumably also explains why the broad-band channel is only marginally
more sensitive than its component bands, as demonstrated by the fact that only 
$\sim 11\%$ of the sources were detected solely in the broad band. 
\cite{sci293} have noted a similar lack of overlap between the
soft and hard source populations detected in deep \chan observations 
of the Galactic plane.

\begin{table}
\begin{center}

\caption{Summary of source detections in each camera and energy band. 
\label{tab:detect}}

\vspace{0.1in}
\begin{tabular}{lcccccr} \hline

& \multicolumn{3}{c}{Energy Band} \\

Camera & Soft & Hard  & Broad & H\&S ${}^a$ & B-only${}^b$ & Total \\ \hline

pn & 90 & 128 & 171 & 22 & 26 & 222 \\

MOS & 135 & 215 & 266 & 43 & 38 & 345 \\

\multicolumn{7}{l}{$^a$ Number of sources detected in {\it both} the
hard and soft bands.} \\
\multicolumn{7}{l}{$^b$ Number of sources detected {\it only} in the broad
band.} \\

\end{tabular}
\end{center}
\end{table}

\subsection{The spectral properties of the XGPS-I sources}
\label{sec:sospec}

The range of spectral hardness exhibited by the XGPS-I sources is 
illustrated in Fig. \ref{fig:hardflux} which shows HR versus
MOS count rate for sources detected in the MOS cameras.
There is clearly a huge spread encompassing the full range
of the HR parameter ({\it i.e.,} $HR = -1$ to $+1$). Given this
scatter, it is not surprising that there is little evidence for
a variation of the average HR with decreasing count rate (as
might be predicted, for example, if fainter sources are on average 
more distant and as a consequence are more strongly absorbed).

\begin{figure}
\begin{center}
\includegraphics[height=3in,width=!,angle=270]{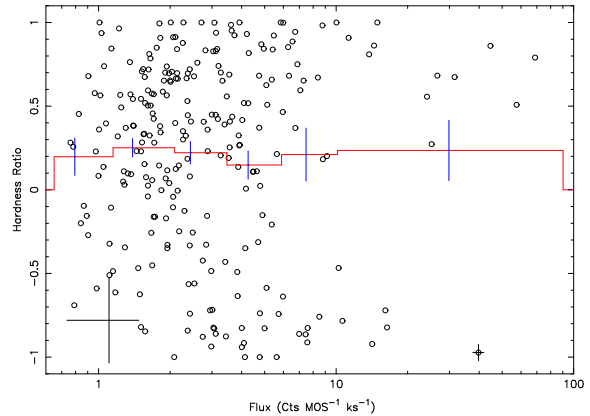}
\caption{The relationship between spectral hardness and source count rate 
for sources detected in the MOS cameras. Error bars are shown for 
representative high and low count-rate points. The histogram charts the 
variation in the average HR with count rate.   
\label{fig:hardflux}}
\end{center}
\end{figure}

We have investigated the typical spectral form  of the XGPS-I source
population by considering the integrated spectra of different subsets 
of sources. In fact, we split the 
population into three groups depending on the HR parameter as follows:
(i) soft sources with HR $< -0.5$; (ii) mid-range sources with
$-0.5 \le $ HR $ \le 0.5$ and (iii) hard sources with  HR $> 0.5$. 
We use the procedure described in \S \ref{sec:data:spec} to extract
the integrated spectra for the individual fields and then sum over
the set of observations to obtain the average spectrum for each
source group. To avoid undue bias, an extremely bright source detected 
in XGPS 9  (XGPS-I J182833-103659 - see \S \ref{sec:trans}), which 
contains a comparable number of counts to all the other sources put together 
was excluded from this process.   

The integrated MOS spectra obtained as above were analysed 
using the XSPEC software package. Following standard practice, the spectra 
were binned prior to analysis  to give a minimum of 20 counts per 
spectral channel. From Fig. \ref{fig:sospec} it is immediately evident 
that the spectra of the three groups of sources are very different.

Initially we fitted the soft-source spectrum in the 0.4--6.0 keV range with 
three different models: a power-law, bremsstrahlung and Mekal thermal 
plasma model, including absorption in each case. The pure continuum models 
provided the best-fits (albeit with
modest reduced $\chi^{2}$) with the 
power-law model requiring a very steep spectral index and the 
bremsstrahlung model requiring a relatively low
temperature (see Table \ref{tab:soft}).  In contrast,
a single temperature solar-abundance Mekal model provided a poor fit to the 
spectrum. However, since the pure continuum models are probably not physically 
realistic characterisations of this soft spectrum, we also investigated
a two-component solar-abundance Mekal model (plus absorption). The result, 
with the two temperature parameters fixed at representative values
(here we use kT=0.25 and 1.5 keV respectively), was a slight improvement 
in terms of $\chi^{2}$ to those obtained for the 
power-law and bremsstrahlung models. Table \ref{tab:soft}
provides details of the fit and Fig. \ref{fig:sospec} compares the 
best-fitting 2-temperature model with the data.

\begin{table}
\begin{center}
\caption{Modelling of the soft-source spectral data
\label{tab:soft}}
\vspace{0.1in}
\begin{tabular}{lccccc} \hline
Model & $N_H {}^a$ & $\Gamma$ & kT$_{1}$ $^b$ & kT$_{2}$ $^b$ & $\chi^2$(dof)\\
\hline
Power-Law & $0.65^{+0.08}_{-0.05}$ & $5.8^{+0.06}_{-0.03}$ &-&-& 164 (107) \\
Brems     & $0.32^{+0.13}_{-0.04}$ &-& $0.38^{+0.03}_{-0.02}$  &-& 173 (106) \\
1-Mekal &  $\sim0.60$            &-& $\sim0.5$              &-& 299 (107) \\
2-Mekal & $0.44^{+0.03}_{-0.03}$ &-& 0.25$^c$  & 1.5$^c$ & 160 (107) \\
\hline
\multicolumn{6}{l}{$^a$ In units of $10^{22} \rm~{cm^{-2}}$.} \\
\multicolumn{6}{l}{$^b$ In keV} \\
\multicolumn{6}{l}{$^c$ Fixed parameter} \\
\end{tabular}
\end{center}
\end{table}

An initial investigation of both the mid-range and hard-source spectra 
(over the spectral range 0.4--8 keV) demonstrated that a simple power-law 
continuum plus absorption model provided a good description of
both datasets with a fairly similar value for the spectral index 
($\Gamma \approx 1.6$) but with the absorption column density for the 
mid-range sample significantly lower than for the hard-spectrum sources.
On this basis, we fitted the two spectra {\it simultaneously} with 
the absorbed power-law model, but with the spectral index as the only tied
parameter. The result was a good fit ($\chi^2=440$ for
421 dof) with $\Gamma = 1.60^{+0.10}_{-0.13}$ and $N_H$ values of 
$0.5^{+0.08}_{-0.08}  \times 10^{22} \rm~cm^{-2}$ and 
$3.7^{+0.3}_{-0.4}  \times 10^{22} \rm~cm^{-2}$ for the
mid-range and hard-source spectra respectively. 
With $\Gamma$ fixed at 1.7 the respective $N_H$ values became 
$0.6^{+0.06}_{-0.05} \times 10^{22} \rm~cm^{-2}$ and 
$3.9^{+0.2}_{-0.3} \times 10^{22} \rm~cm^{-2}$.  
A comparison of the best-fitting models 
with the data are again shown in Fig. \ref{fig:sospec}.

Both the mid-range and hard source spectra contain a line emission feature 
in the 6--7 keV band, consistent with Fe K$_{\alpha}$ emission.  Although 
the data are of limited quality, we determine the line centroid values to be 
$ 6.59 \pm 0.07$ keV and $ 6.88 \pm 0.06$ keV for the medium and hard sources 
respectively; the equivalent widths are measured to be $370\pm250$ and 
$240\pm110$ eV.

The fact that the soft-source spectrum is well fitted by
a canonical 2-temperature model with a relatively low
absorption column is consistent with the bulk of the 
soft population being relatively nearby active stars. 
The spectra of the mid-range and hard-source samples are less
easy to characterise. Certainly many of the faint hard sources
may be AGN (see \S \ref{sec:pops}) but the relatively hard continuum spectrum 
and iron-line properties also match the spectral properties of cataclysmic 
variables (CVs) and RS CVns. For example, CVs often exhibit a
two-temperature thermal spectrum with $kT \sim 0.5-1$ keV and 
$\sim 5-10$ keV (eg. Baskill et al. 2003, submitted). With significant 
line of sight absorption the latter
component dominates and readily mimics the hard power-law form inferred above.

\begin{figure}
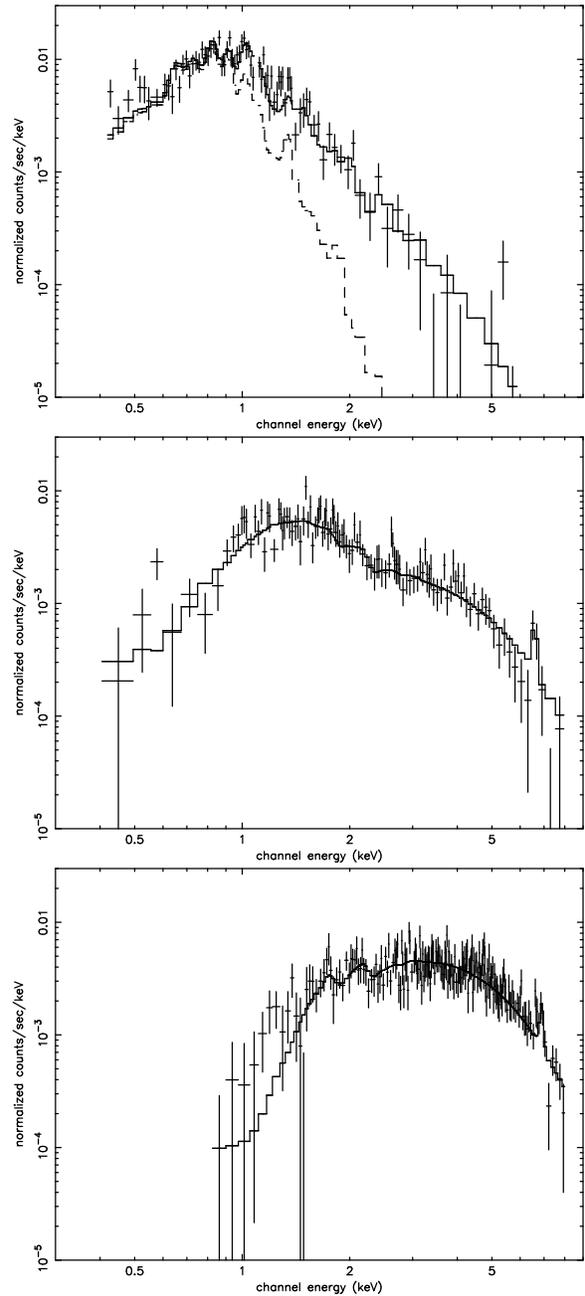

\begin{center}
\includegraphics[height=3in,width=!,angle=270]{gps_fig7a.ps}
\includegraphics[height=3in,width=!,angle=270]{gps_fig7b.ps}
\includegraphics[height=3in,width=!,angle=270]{gps_fig7c.ps}
\caption{{\it Top Panel:} The integrated EPIC MOS spectrum of the soft
XGPS-I sources with HR $< -0.5$.  {\it Middle Panel:} The integrated EPIC 
MOS spectrum of the XGPS-I sources with mid-range hardness ratios
{\it i.e.,} $-0.5 \le $ HR $ \le 0.5$.
{\it Bottom Panel:} The integrated EPIC MOS spectrum of the hard XGPS-I 
sources with HR $ > 0.5$. In each case the histogram represents the 
besting-fitting spectral model described in the text. In the top panel
we also show the contribution of the lower temperature Mekal component
to the measured spectrum.
\label{fig:sospec}}
\end{center}
\end{figure}

\subsection{The XGPS-I Source Counts}

In order to study the number density of discrete X-ray sources as a function 
of count rate it is necessary to correct for the variation in the source 
detection sensitivity across the set of XMM fields which comprise
the survey. Here we concentrate solely on the sources detected in 
the MOS cameras. 

The first step in the correction process was to calculate a sensitivity map 
for source detection (in ``on-axis'' count-rate units) for each XGPS-I 
observation 
based on the exposure map (which accounts for vignetting and other
relevant factors such as chip gaps) and the derived MOS background map. 
The total survey area over which a source of a given count rate was detectable
was then readily calculated by summing over the set of sensitivity
maps comprising the XGPS-I survey. The derived effective area curves
are shown in Fig. \ref{fig:sens} for the three energy  bands of the survey.

\begin{figure}
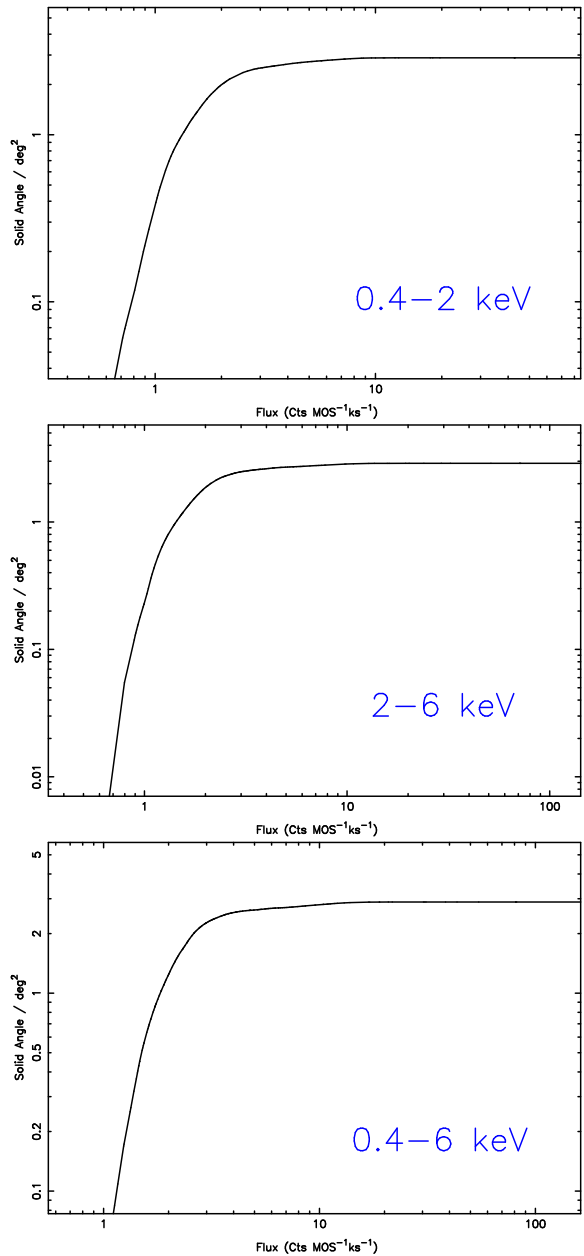

\begin{center}
\includegraphics[height=3in,width=!,angle=270]{gps_fig8a.ps}
\includegraphics[height=3in,width=!,angle=270]{gps_fig8b.ps}
\includegraphics[height=3in,width=!,angle=270]{gps_fig8c.ps}
\caption{Sensitivity curves for the source count analysis based on
detections with the MOS cameras in three 
energy bands. The total solid angle coverage of the survey (for very
bright sources) is $\sim 3$ square degrees.
\label{fig:sens}}
\end{center}
\end{figure}

The X-ray source counts are then constructed by summing the contributions
of individual sources after correction for the survey sensitivity. 
For example, consider a source detected at some particular offset angle 
in one of the XGPS-I observations. Its on-axis count rate is obtained by 
simply dividing the corrected net counts by the value 
of the exposure map at the source position. We then use the derived
sensitivity curves to determine the solid angle ($\Omega$) over which a source
of that count rate was detectable. This source then contributes $1/\Omega$
to the source counts at its measured count-rate value. The final source
count is obtained by summing the contributions of all the detected
sources\footnote{Sources with  $quality < 0.8$, such as those
located at a CCD chip edge or strongly affected by bad pixels were excluded.}.

In order to obtain an estimate of the magnitude of the error that
should be assigned to the derived source counts at a given flux 
we have carried out a Monte Carlo simulation of the {\it post-detection}
process used to construct the source counts ({\it nb.} bearing in mind that
with integral counts the measurements are not independent from point to point).
This simulation also demonstrated that the changing gradient of the 
sensitivity curves at low fluxes introduces a significant bias in the 
source counts; in effect the Poissonian variation in the measured flux of
a source has an asymmetric effect on the value of $\Omega$ that is derived.  
We correct for this bias by simulating the source counts both with and 
without such flux errors, noting the differences and adjusting the 
measured data accordingly.  In  practice this procedure resulted in 
a reduction in the inferred number density of sources at the survey limit
by up to 40 per cent.

Fig. \ref{fig:logn} shows the corrected integral source counts
in the three bands. It is evident that XGPS-I survey detects discrete
X-ray sources in the Galactic plane down to surface density of roughly
200 per square degree.

By linearly fitting the data in binned, differential form, we determined 
the slope of the integral counts to be $-1.5 \pm0.2$ for both the soft and 
hard sources and $-1.3 \pm0.2$ for the broad band sources. (These values 
represent the slopes of the source counts after excluding, in each case,
a handful sources at the bright end of the flux range.)

\begin{figure}
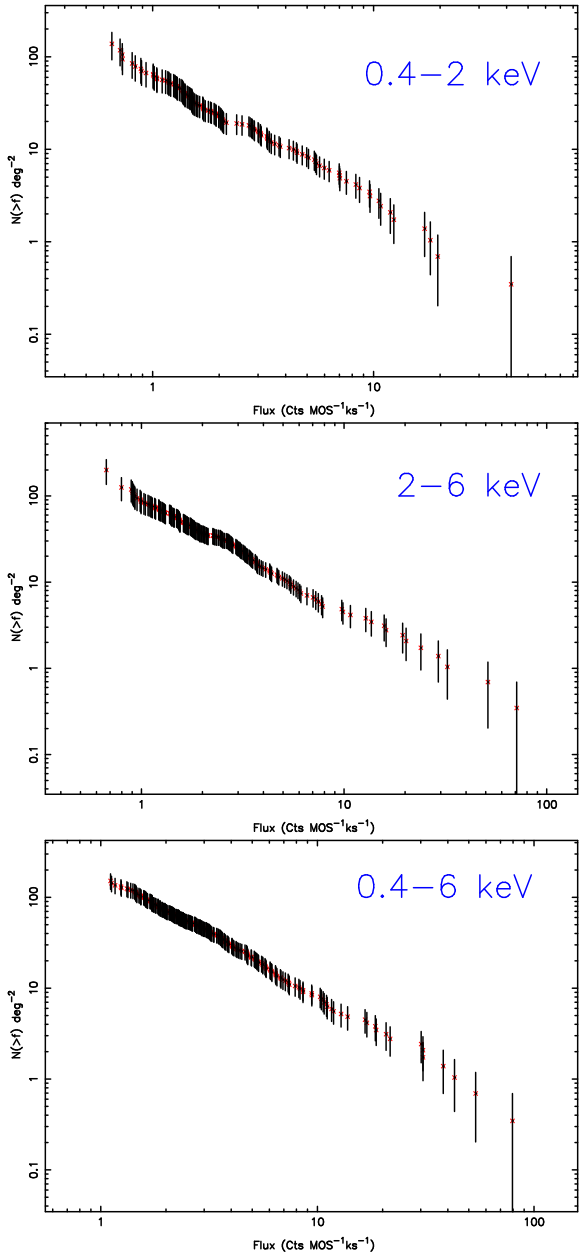

\begin{center}
\includegraphics[height=3in,width=!,angle=270]{gps_fig9a.ps}
\includegraphics[height=3in,width=!,angle=270]{gps_fig9b.ps}
\includegraphics[height=3in,width=!,angle=270]{gps_fig9c.ps}
\caption{The derived X-ray source counts plotted in three energy bands and 
corrected at faint fluxes for the bias in the coverage correction
induced by flux errors (see text).
\label{fig:logn}}
\end{center}
\end{figure}

\subsection{Optical/X-ray Source Correlations}

Although the X-ray positions typically have statistical 
errors of $ \le 4''$ (see \S \ref{sec:detect}), we have searched for optical 
counterparts within
a nominal $6''$ error circle. Specifically we have used optical data
from the SuperCOSMOS digitisation of the sky survey plates from the 
UK Schmidt telescope (UKST). Appendix A identifies the brightest optical 
source (if any) on the red (R) plate within the error circle of each XGPS-I 
source and quotes the corresponding optical R magnitude. Cross references to 
the optical  source in the USNO-A2.0 catalogue and/or the SIMBAD database
are also noted. Of the 424 X-ray point sources, 188 have possible 
optical counterparts identified by this procedure.

The correlation of optical magnitude versus X-ray count rate 
(here we focus on detections with $quality > 0.8$ in the MOS cameras) 
is a scatter diagram.  Similarly a plot of X-ray hardness ratio versus 
optical R magnitude also shows 
significant scatter (Fig. \ref{fig:hardmag}), although there is hint
of X-ray spectral hardening as one goes to optically fainter sources
in the range R $=$ 12-18.

\begin{figure}
\begin{center}
\includegraphics[height=3in,width=!,angle=270]{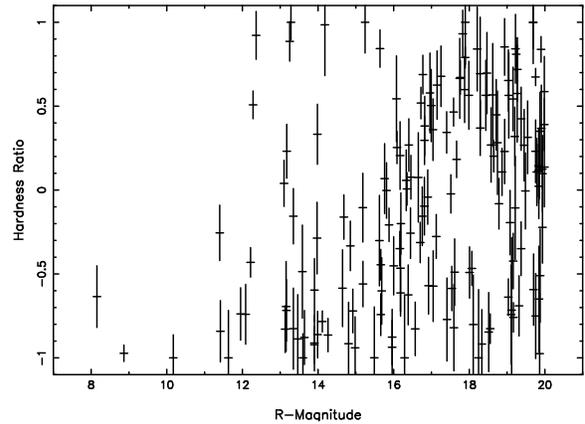}
\caption{X-ray spectral hardness ratio versus optical R magnitude. 
This refers to the brightest optical object in the X-ray error 
circle of each XGPS-I source. The probability that the
optical source is the true optical counterpart of the X-ray source 
decreases both with increasing R and hardness ratio - see the discussion 
in the text.
\label{fig:hardmag}}
\end{center}
\end{figure}

We have investigated how the number of optical/X-ray correlations
varies with optical magnitude for three subsets of sources divided
according to the  X-ray hardness ratio ({\it i.e.,} the soft-, mid-range and 
hard-spectrum samples defined earlier). Fig. \ref{fig:fractopt} shows how the 
fraction of X-ray sources with an associated optical source rises
with increasing R. For the hard X-ray source 
sample, the rate of optical correlation is essentially the same
as the chance rate. However, both the soft and mid-range samples have
significantly higher rates of optical/X-ray associations than
expected by chance. For these, we can compare the observed
and chance rates to estimate the fraction of genuine optical identifications
within the full list of optical associations (see Fig. \ref{fig:fractopt},
{\it lower panel}).

\begin{figure}
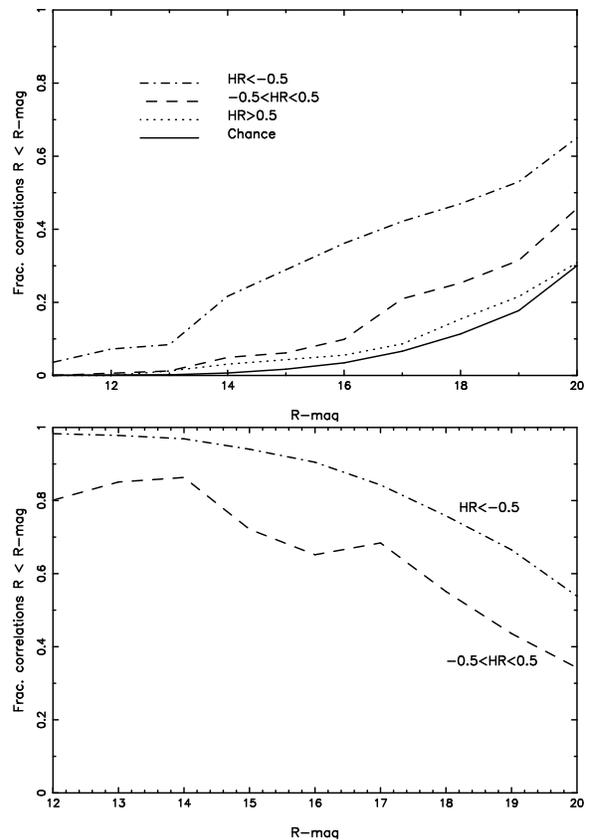

\begin{center}
\includegraphics[height=3in,width=!,angle=270]{gps_fig11a.ps}
\includegraphics[height=3in,width=!,angle=270]{gps_fig11b.ps}
\caption{{\it Top Panel:} The fraction of X-ray sources with an associated 
optical source plotted versus the limiting R magnitude of the optical sample.
The three upper curves correspond to X-ray sources with
spectral hardness in three ranges, namely soft sources with HR $< -0.5$,
mid-range sources with $-0.5 \le $ HR $ \le 0.5$ and hard sources with  
HR $> 0.5$. The lower curve shows the chance coincidence rate for
finding an optical source in a $6''$ error circle in this region of the sky.
{\it Bottom Panel:} The fraction of the optical/X-ray associations that are
likely to represent real identifications. The two curves correspond
to the soft (upper) and mid-range (lower) spectral samples.
\label{fig:fractopt}}
\end{center}
\end{figure}

The X-ray sources with soft spectra (HR $< -0.5$) have a particularly
high rate of association with bright optical objects. For example,
$\sim 45$ per cent of such sources have an optical object brighter than R=18 
within $6''$ of the X-ray position  and of these roughly 75 per cent 
are likely to be the correct counterpart. At R=20 the two factors become 
$\sim 65 $ and $\sim 55 $ per cent respectively.  On the basis of the inferred 
X-ray/optical
ratio and the X-ray spectral characteristics discussed earlier,
it is likely that many of these soft X-ray sources are nearby late-type stars
with active coronae. 

Having identified a subset of the optical/X-ray associations which have
a relatively high probability of being the correct identification, we can
use the measured optical to X-ray positional offsets to check the
astrometry of the X-ray positions, including any component
relating to an overall shift (and rotation) of the 
\xmm reference frame.  Fig. \ref{fig:astrom} shows the radial distribution of
the optical/X-ray offsets for the soft sources with associated 
optical objects brighter than R=20. Allowing for a uniform distribution of 
chance
coincidences we find the radius encompassing 68 per cent (90 per cent) of 
the ``real''
identifications is $3.3''$ ($4.7''$) which is comparable to
our earlier estimate of the statistical errors associated with the 
X-ray positions.  This analysis demonstrates that any systematic astrometric
shift of the \xmm reference frame (for each field) to the true celestial 
frame must be small (of the order $1-2''$ at most), a result that is in 
accord with other studies
({\it e.g.,} \citeb{aa382x}; \citeb{an324x}).
Of the 22 XGPS fields, 17 have at least one
soft source with an optical counterpart brighter than R=20
within $4.7''$. Conversely we calculate that an incidence of 5 fields with 
zero correlations is not a particularly unlikely event. Unfortunately this 
does mean that for the latter fields we have no independent check
of the \xmm aspect solution (the fields in question
are Ridge 4 and XGPS 2,4, 6 \& 14), but we have no reason to believe the
astrometric accuracy of any of these fields is anomalous.

\begin{figure}
\begin{center}
\includegraphics[height=3in,width=!,angle=270]{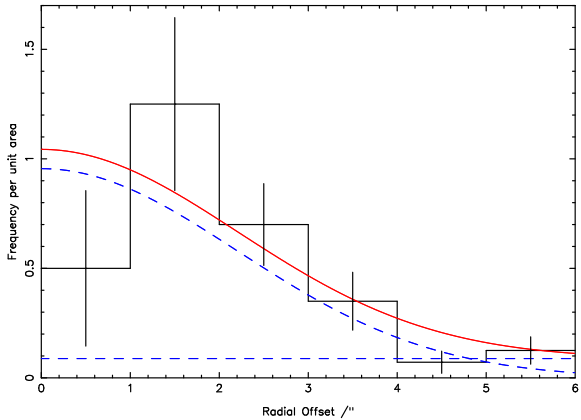}
\caption{The distribution of the optical/X-ray offsets measured
for the soft source sample in radial bins of $ 1''$ width. The error bars 
correspond to $\pm1 \sigma$.  The solid-line represents the sum of the
uniform distribution of the chance coincidences and the assumed
Gaussian distribution of the real identifications. The latter has
$\sigma=2.2''$ corresponding to a 68 per cent (90 per cent) probability
error circle radius of $3.3''$ ($4.7''$). 
\label{fig:astrom}}
\end{center}
\end{figure}

\subsection{A Bright Transient Source}
\label{sec:trans}

Only one source in the XGPS-I catalogue is bright enough to merit individual 
spectral extraction.  The source, designated  XGPS-I J182833-103659, is 
located 
at RA, Dec (J2000) $\rm 18^h~28^m~34.0^s$ , -10\deg $36' 59''$ (Galactic
 coordinates $l=$20.9\deg, $b=$0.2\deg).  Fitting an absorbed thermal 
bremsstrahlung model to the measured spectrum yields a temperature of 
$\sim 7$ keV and an absorption column of $\sim 5 \times 10^{22} 
\rm~{cm^{-2}}$ (see Fig. \ref{fig:trans}). This column density is consistent 
with either an extragalactic or a distant Galactic origin.  
In the latter case (assuming a distance of $\sim 15$ kpc) the observed 
flux is equivalent to an X-ray luminosity of $\sim 10^{35} \rm~{erg} 
\rm~{s^{-1}}$.  

\cite{aa392} discovered 6 type I X-ray bursters in \sax Wide Field Camera (WFC)
observations, one of which is positionally coincident with this bright XGPS-I
source, although the WFC 99 per cent confidence error circle of $2.8'$ 
is relatively large.
The peak flux measured by \sax for this source (during a burst) was 
$(1.1 \pm0.4) \times 10^{-8}$ \ergseccm (2--10 keV) with a burst duration
of $\sim$ 30 seconds. In the same 
energy band we measure $\sim~7.5~\times~10^{-12}$ \ergseccm (after correcting 
for absorption). This is more than three orders of magnitude fainter than
the burst peak, but may be only a factor $\sim10$ fainter
than the high-state persistent flux of this source 
for which \cite{aa392} quote only an
upper limit of $< 1.9 \times 10^{-10}$ \ergseccm.
XGPS-I J182833-103659 shows no variation in its light curve over the
short \xmm observation,  indicating that we are detecting
persistent emission.
The source position was in fact  covered by two XGPS-I observations,
XGPS 9 and Ridge 5, but the source was detected only in the former.  
This places an upper flux limit on the
low state of the source of approximately $2~\times~10^{-14}$ \ergseccm~
in the 2--10 keV band. This source thus clearly shows significant
variability: by a factor $\sim 300$ between \xmm observations and quite
possibly by a much larger factor overall, making it very likely to be a
previously unrecognised X-ray
transient source. X-ray bursts are of course 
commonly associated with X-ray transient
systems. The observed low state is also consistent with
the absence of the source from the catalogue derived from
the \asca Galactic Plane Survey (\citeb{a134}).

\begin{figure}
\begin{center}
\includegraphics[height=3in,width=!,angle=270]{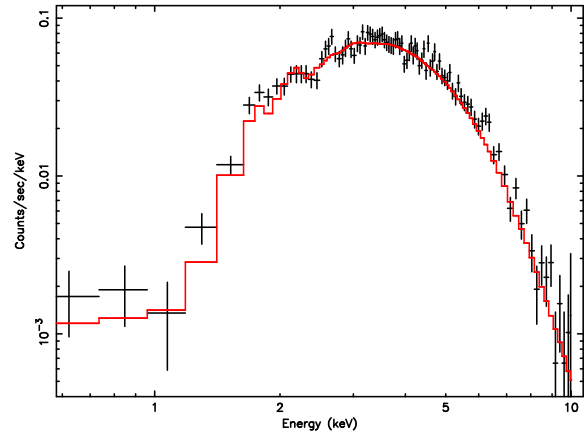}
\caption{The measured count-rate spectrum of a bright transient source.
The histogram corresponds to the best-fitting model detailed in the
text.
\label{fig:trans}}
\end{center}
\end{figure}

\subsection{Comparison with \asca}
The region surveyed by XGPS-I is entirely covered by the \asca Galactic Plane 
Survey (\citeb{a134}) which resolved 163 sources in the Galactic Plane within a
longitude span of 90\deg  centred on $l=0$\deg.  Of these 163 
sources nine fall within the nominal region covered by the XGPS-I 
observations. We have detected seven of these sources 
as summarised in Table \ref{tab:asca}. 
Note that, two of the \asca sources are linked to the same 
XGPS-I source as a consequence of the relatively poor spatial resolution
and large positional errors of the former.

\begin{table}
\begin{center}
\caption{Correlation of XGPS-I sources with the \asca catalogue. 
\label{tab:asca}}
\vspace{0.1in}
\begin{tabular}{ll} \hline

XMM ID   & \asca Name \\ \hline

XGPS-I J182534-121454 & AX J182538-1214 \\
%(R1\5)

XGPS-I J182846-111711 & AX J182846-1116 \\
%(R4\_14)

XGPS-I J182525-114525 & AX J182530-1144 \\
%(GPS1\_12)

XGPS-I J183038-100249 & AX J183039-1002 \\
%(GPS14\_4)

% GPS17\_8 & AX J183114-0943 \\

XGPS-I J183117-100921 & AX J183116-1008 \\
%(GPS16\_2)

XGPS-I J183209-093906 & AX J183206-0938 / AX J183206-0940 \\
%(GPS17\_14)

- & AX J183114-0943 \\
- & AX J182651-1206 \\

\hline

\end{tabular}
\end{center}
\end{table}

\section{Discussion}

\subsection{Source Populations and the log N - log S Relation}
\label{sec:pops}

The source number versus flux (log N - log S) relation can provide important 
information on the spatial distribution and the luminosity functions of the 
various Galactic source populations.  Here we combine our present 
measurements with those from other missions to examine how various 
categories of source may contribute to the observed hard band source counts. 
An investigation of the soft band log N - log S including a comparison
with earlier {\it ROSAT} measurements  will be the subject of a 
future paper.

In order to relate the \xmm measurements to observations from other
satellites it is necessary to convert the measured source counts from 
count-rate to flux units. Table \ref{tab:conv} lists the conversion 
factors from MOS count rate in the 2--6 keV band to the corresponding flux 
(in $\rm~erg~s^{-1}~cm^{-2}$) in the 2--10 keV band for a variety of spectral 
forms calculated using PIMMS (\citeb{leg3}).   

In the present analysis we adopt a factor $2.6 \times 10^{-14}$
$\rm~erg~s^{-1}~cm^{-2}$ / MOS count ks$^{-1}$ corresponding to 
a power-law source spectrum with spectral index $\Gamma = 1.7$ absorbed
by a column density $N_H = 1 \times 10^{22} \rm~cm^{-2}$.  This is clearly 
a compromise given the range of spectral form established
earlier (see \S \ref{sec:sospec}); we estimate that the effective uncertainty 
in the flux scaling may be as large as $\pm 30\%$. 

\begin{table}
\begin{center}
\caption{Factors$^a$ to convert from the MOS count rate in the 2--6 keV band 
to the observed flux in the 2--10 keV band for different power-law spectra 
with a range of interstellar absorptions. 
\label{tab:conv}}
\vspace{0.1in}
\begin{tabular}{lccccc} \hline
 $\Gamma \backslash N_H {}^b$ ~~~~~~~~~~~~~~~~~& 0.0 & 1.0 & 3.0 & 5.0 & 7.0 \\ 
\hline
1.4 & 2.7 & 3.0 & 3.4 & 3.9 & 4.4 \\
1.7 & 2.4 & 2.6 & 3.0 & 3.4 & 3.9 \\
2.0 & 2.1 & 2.3 & 2.7 & 3.1 & 3.4 \\
\hline
\multicolumn{6}{l}{$^a$ In units of $10^{-14} \rm~erg~s^{-1}~cm^{-2}$ / 
(MOS count $\rm~ks^{-1}$)} \\
\multicolumn{6}{l}{$^b$ In units of $10^{22} \rm~cm^{-2}$.} \\
\end{tabular}
\end{center}
\end{table}

Fig. \ref{fig:all_data} shows the measured log N - log S relation
in the 2--10 keV band based on the XGPS-I measurements. Here we 
have clipped one high flux point with large error bars and two low-flux 
points requiring very large coverage correction, from the equivalent 
representation in Fig. \ref{fig:logn}. For comparison the source counts
derived from the extensive survey of the Galactic Plane between
$l= \pm$45\deg carried out by 
\asca (\citeb{a134}) and from recent deep \chan observations
in the Galactic Plane at $l=$28\deg (\citeb{sci293}) are also shown. 

As can be seen from Fig. \ref{fig:all_data}, the flux range probed by 
the XGPS-I measurements is intermediate between that sampled by the
\asca and \chan programmes. The agreement between the XGPS-I and \asca
surveys is rather good given the very different coverage of the Galactic 
Plane inherent in the two programmes. 
The agreement between the XGPS-I and \chan source counts is also good 
bearing in mind the different pointing directions and the fact that at
$\sim 3 \times 10^{-14} \rm~erg~s^{-1}~cm^{-2}$ there are only $\sim 6$
sources in the latter survey.  Based  on this compilation
the log N - log S relation appears to first flatten, then steepen, then 
flatten again as one moves from bright sources at $\sim 10^{-10} 
\rm~erg~s^{-1}~cm^{-2}$ to faint sources at a limiting flux of
$\sim 3 \times 10^{-15} \rm~erg~s^{-1}~cm^{-2}$ in the 2--10 keV band.

\begin{figure}
\begin{center}
\includegraphics[height=3in,width=!,angle=270]{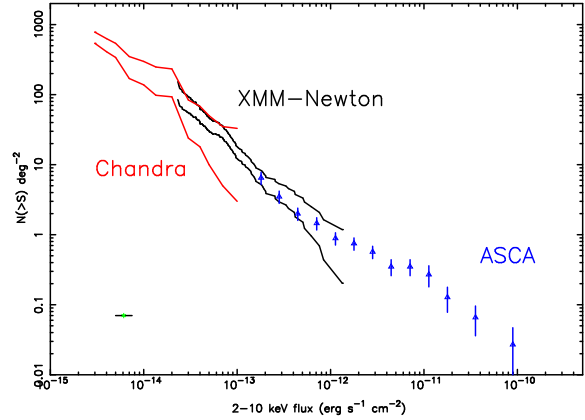}
\caption{The 2--10 keV log N - log S relation measured in the Galactic Plane
based on {\it ASCA}, \xmm and  \chan observations (see the text for 
references). In the case of both the \xmm
and \chan measurements we show the upper and lower bounds of the derived 
source counts rather than individual data points. The magnitude 
of the relative flux scaling uncertainty applicable to the three data sets 
is indicated by the horizontal error bar. \label{fig:all_data}}
\end{center}
\end{figure}

The first step in modelling the measured composite log N - log S
relation in terms of various underlying source populations is to
quantify the contribution of extragalactic sources. It has in fact
been recently demonstrated that even in heavily obscured regions of the 
Galactic 
plane the X-ray source counts measured at faint fluxes in the hard
band are dominated by this component (\citeb{sci293}).
The extragalactic log N - log S relation in the 2--10 keV 
band has been determined over a wide range of X-ray flux from HEAO-1 A2 
observations at the bright end (\citeb{a253}) through to recent ultra-deep 
\chan observations which probe below $\sim 10^{-15}$ \ergseccm 
(\citeb{a566a}; \citeb{a566b}; \citeb{aj126b}; \citeb{a588};
\citeb{a600}).
%(\cite{n404}, \cite{a551}).  
Here we use the empirical form for the extragalactic 2--10 keV log N - log S
specified by \cite{a560} based on a comparison of deep
\chan observations with {\it ASCA}, {\it BeppoSAX} and other data sets. The
integral form of the extragalactic source counts flatten from a power-law slope
of $-1.67$ at intermediate fluxes to a value of $-0.58$ at faint fluxes with 
the break occurring near $\sim 2 \times 10^{-14}$ \ergseccm. 

A very important factor in modelling the extragalactic contribution 
to the Galactic Plane log N - log S is the signal loss
due to absorption in the line-of-sight column density through
Galaxy. Our spectra analysis of the sources with the hardest spectra (see \S 
\ref{sec:sospec}) sets a {\it lower limit} of $N_H = 3.9 \times 10^{22} 
\rm~cm^{-2}$ for an assumed power-law source spectrum with $\Gamma = 1.7$,
whereas the bright transient source discussed earlier (\S \ref{sec:trans})
required $N_H = 5.0 \times 10^{22} \rm~cm^{-2}$.
By way of comparison, \cite{sci293} argue that  $N_H = 4 - 6 \times 
10^{22} \rm~cm^{-2}$ at $l=28$\deg when one accounts for both neutral and 
molecular hydrogen along the line of sight.  Based on the HI measurements of
\cite{ara28} and molecular hydrogen measurements of \cite{a547} one might 
infer a similar value for the XGPS-I region.  
On the other hand \cite{aa374} measure a foreground Galactic 
column density of $7.9\pm 0.5 \times 10^{22} \rm~cm^{-2}$ for a cluster of 
galaxies at (l,b) = (21.3,-0.7).  Here we adopted a hard band transmission 
factor of 0.68 corresponding to a line of sight column density of 
$5 \times 10^{22} \rm~cm^{-2}$ (for a power-law $\Gamma=1.7$ source spectrum)
which, in broad terms, aligns the extragalactic prediction with the 
observed \chan source counts at faint fluxes (see Fig. \ref{fig:logn_h_fin}).
Clearly variation in the 
Galactic $N_H$ from field to field in the Galactic Plane will introduce 
a significant variance in the extragalactic contamination of the log N - 
log S relation; of necessity here we present only an approximate
description of a very complicated situation. 

We have investigated the possible contribution of various Galactic source 
populations to the measured composite log N - log S relation through the use 
of relatively simple prescriptions for the source luminosity  function, the 
source distribution in the Galaxy and the effects of absorption.  In brief, 
the predicted source counts are calculated by a numerical integration 
along a line of sight at (l,b) = (20,0). We assume the maximum diameter of 
the Galaxy is 20 kpc and the Galactocentric radius of the Sun is 8.5 kpc. 
We model the absorption in the plane in terms of a local
hydrogen density of 0.55 $\rm~cm^{-3}$.
We assume the source and particle densities decline exponentially with 
respect to Galactocentric radius ($R$) and height above the plane ($z$)
(the assumed scale factors were 8500 kpc and 200 pc in $R$ and $z$ 
for the sources and 8500 kpc and 100 pc for the particle density).

We first consider relatively luminous Galactic X-ray binary sources containing 
either a neutron star or (in a few cases) a stellar mass black-hole.
Low-mass X-ray binaries (LMXBs) are found preferentially in the Galactic 
Bulge and Galactic Centre regions whereas high-mass X-ray binaries (HMXRB)
tend to avoid the inner 3-4 kpc of the Galaxy but are widely distributed 
in the Galactic disk (\citeb{aa391}). The composite log N - log S measured
at $l=20$\deg might therefore include contributions from both populations. 
In order to model the combined LMXB/HMXRB contribution we assume a 
power-law form for luminosity function with a slope of $-1.3$ in the 
differential form (cf. \citeb{aa391}).  In practice, a luminosity function 
restricted to the range $10^{34}-10^{36}$ \ergsec proved sufficient
to account for the observed form of the log N - log S relation at
the bright source end (see Fig. \ref{fig:logn_h_fin}).
The normalisation of the binary luminosity function needed to
match the log N - log S relation translates via the source distribution 
model to a Galactic population of $\sim 200$ such X-ray binaries with 
an integrated Galactic X-ray luminosity of $\sim 1.6 \times 10^{37}$ \ergsec.

With the bright and faint ends of the measured log N - log S relation
represented respectively by Galactic X-ray binaries and the breakthrough
of extragalactic sources, an excess number of sources (relatively
to the prediction) is most apparent in the flux range $10^{-13}$ to
$10^{-12}$ \ergseccm. The requirement on any source population invoked
to fill this gap is that its source count must be relatively steep at the 
top end of this range but should gradually turn over below $10^{-13}$
\ergseccm, so as not to overpredict the total source density in the 
flux range sampled by the \chan observations. 

For illustrative purposes
we consider a source population with an X-ray luminosity function 
described by a log-normal function centred on $L_X = 10^{31}$ \ergsec 
with $\sigma=1.0$ and a local spatial density of $\sim 10^{-6} \rm~ pc^{-3}$.
A source with $L_X = 10^{31}$ \ergsec, at a distance of 1 kpc has an X-ray 
flux of $10^{-13}$ \ergseccm.  At larger distances (and hence lower fluxes), 
the effects of increasing absorption will serve to 
flatten the counts of such sources.  In addition by $ \sim 10^{-14}$ 
\ergseccm~the most luminous sources in the population are detectable 
out to the edge of the Galaxy with the result that the overall log N - log S 
relation flattens further.  Fig. \ref{fig:logn_h_fin} shows the
predicted source count relation for the low-luminosity source population
considered above.  In combination the two Galactic source populations plus 
the extragalactic
component provide an excellent match to the observed composite 
log N - log S curve.

What class of X-ray source might comprise the low-luminosity
population considered above?  The most likely candidate population
is cataclysmic variables (CVs), close binary systems in which a white dwarf 
accretes material from a Roche-lobe filling late-type companion. CVs 
are often relatively bright hard X-ray sources with 
$L_X = 10^{30-32}$ \ergsec ({\it e.g.,} \citeb{aa327}).
In an earlier analysis, \cite{mcv291} 
suggested that X-ray faint CVs might show up in large numbers in deep 
Galactic surveys carried out in the hard X-ray band if their
typical X-ray luminosity is $L_X = 10^{31}$ \ergsec (2--10 keV) and space 
density is $\sim 10^{-5} \rm~ pc^{-3}$. The latter value is compatible with
the CV space density derived empirically by \cite{patt84}   and is 
not out of line with at least some theoretical estimates (eg. \citeb{kolb93}).
In this context, the space density assumed in our modelling of the low-
luminosity population becomes a rather conservative requirement particularly 
since other categories of source, such as RS CVn binaries 
({\it e.g.,} \citeb{aj126a} and references therein) and Be-star X-ray binaries
in quiescence  ({\it e.g.,} \citeb{mn293}),  might also contribute to the 
low-luminosity population. The overall requirement is for a Galactic 
population of $1.2 \times 10^{5}$ objects which produce a
Galactic X-ray luminosity of $1.3 \times 10^{37}$ \ergsec in the 2--10 keV 
band, comparable to the integrated X-ray luminosity inferred earlier
for the X-ray binary population.

\begin{figure*}
\begin{center}
\includegraphics[height=5in,width=!,angle=270]{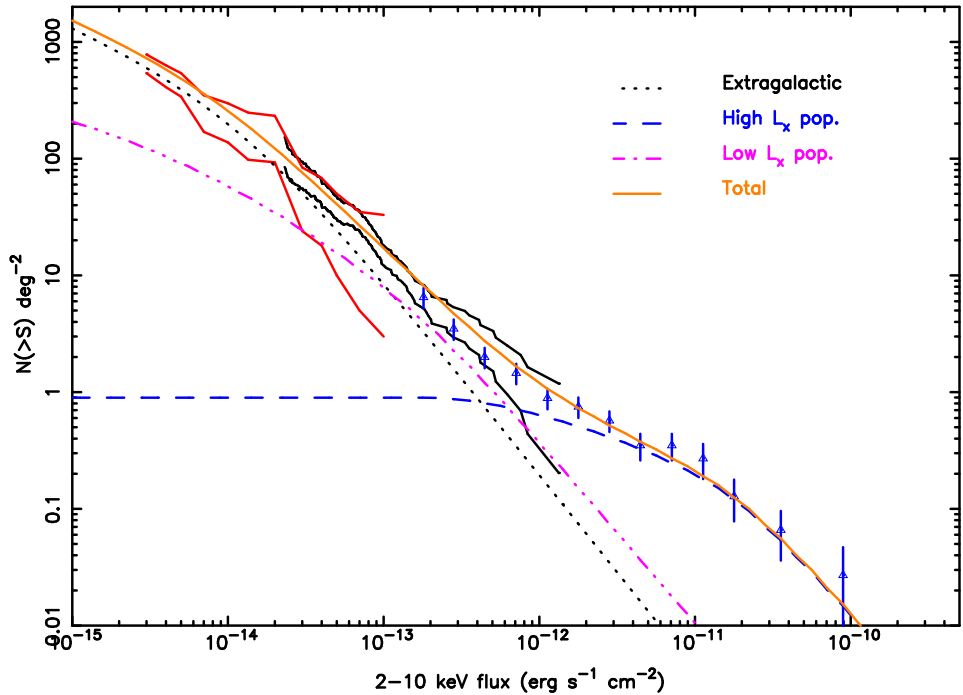}
\caption{The measured 2--10 keV log N - log S relation compared with
with the predicted contributions of various X-ray source populations
(see the text).  \label{fig:logn_h_fin}}
\end{center}
\end{figure*}

\subsection{Contribution of Discrete Sources to the GRXE}

Using all XGPS-I observations except the two most contaminated by flaring 
(XGPS 14, XGPS 15 ) and the one containing the bright transient source
(XGPS 9), we have measured the total full-field MOS count rate (including the 
resolved sources) in the hard \xmm band (2--6 keV).  After correcting for the 
underlying instrumental background ({\it e.g.,} see \citeb{nps}) and for 
mirror vignetting of the sky background signal, we obtain $3.7 \pm 0.1$ 
MOS~\ctsecdeg.  This signal is substantially larger than that measured at 
high Galactic latitude in the MOS cameras consistent with the presence of 
the GRXE in the field of view for all the XGPS-I pointings.

Applying the same count-rate to flux conversion factor as used in \S5.1, the 
{\it total} surface brightness of the GRXE corresponds to $9.6 \times 
10^{-11}$ \ergseccmdeg~in the 2--10 keV band. Earlier estimates put the 
value variously at 11 $\times 10^{-11}$ \ergseccmdeg~ 
({\it Chandra}; \citeb{sci293}), 5.2 $\times 10^{-11}$\ergseccmdeg~ 
({\it ASCA}; \citeb{a134}), and 2.5 $\times 10^{-11}$\ergseccmdeg~ 
({\it RXTE}; \citeb{a505}), depending on the region of sky surveyed.

By integrating the {\it observed} hard band X-ray source counts we find that 
the resolved sources with count rates in the range 0.7 to 70 MOS count 
ks$^{-1}$ (2--6 keV) (on-axis) contribute 0.34 \ctsecdeg, corresponding 
to 9\% of the observed surface brightness. In addition, we estimate that the 
contribution of the residual extragalactic background after transmission 
through Galactic $N_H$ of $5 \times 10^{22} \rm~cm^{-2}$ amounts to a 
further 10 per cent of the measured surface brightness.  Since the 
extrapolation of the log N - Log S curves for the high-luminosity and 
low-luminosity Galactic source populations below the XGPS-I detection 
threshold 
adds little to the integrated signal,  the implication is that $\sim 80$ 
per cent of the measured GRXE surface brightness remains unaccounted for,

The deep \chan observations show that extragalactic sources dominate down 
to  fluxes of $\sim 3 \times 10^{-15}$ \ergseccm~(2--10 keV) (\citeb{sci293}).
A new Galactic population, contributing significantly to the GRXE, might 
emerge at fainter fluxes but the requirement (deduced by scaling the 
properties of the low-luminosity population considered earlier) of, say, 
$L_X = 10^{28}$ \ergsec~combined with a space density of 
$10^{-2} \rm~ pc^{-3}$, does 
not fit any known population of sources.  It would appear therefore that the 
bulk of the GRXE is truly diffuse in origin although the origin
is still uncertain. Possible mechanisms 
include the interaction of low-energy cosmic-ray electrons or ions
with interstellar matter (\citeb{a543b}; \citeb{a543c}), in-situ electron 
acceleration (\citeb{a581a}; \citeb{a581b}) and
magnetic reconnection ({\citeb{a551b}).  This issue will be 
addressed in a later paper on diffuse emission from the XGPS survey region.

\section{Conclusion}

The XGPS-I survey, which covers approximately three square degrees of the
Galactic Plane near $l=20$\deg, has resulted in a catalogue containing 
over 400 discrete X-ray sources. The measured X-ray source counts trace
the source population down to a limiting flux of $\sim 2 \times 10^{-14}$ 
\ergseccm~in the 2--10 keV band at which point the source density is
between 100--200 sources per square degree. Consistent with an earlier
\chan study, the source counts at this flux are predominately
due to extragalactic sources, despite the fact that the fluxes of 
extragalactic objects
are significantly suppressed by absorption in the Galactic plane. However, 
the conclusion of the present work is that at fluxes above 
$10^{-13}$  \ergseccm (2--10 keV) Galactic source populations
do come to the fore. 

The Galactic source population observed between 
$10^{-13}$ and $10^{-12}$ \ergseccm~could comprise largely
CVs and RS CVn systems with X-ray luminosities in the 
range $10^{30-32}$ \ergsec~but the details remain uncertain 
on the basis  of the X-ray information alone.
Extensive programmes to identify and characterise  optical/infra-red 
counterparts are required, although this will be taxing given the
high obscuration and high object density in the Galactic plane.

The present work demonstrates that the strategy of the XGPS programme, 
namely the use of shallow observations to give relatively wide angle 
coverage is close to optimum in terms of maximising the number of
Galactic source detections.

\section{Acknowledgements}

ADPH acknowledges support from PPARC in the form of a research studentship
and DJH acknowledges support from NASA grant NAG5-9870. We are also
very happy to acknowledge the underpinning contributions made by the ESA 
Science Operations (SOC) team and the \xmm~EPIC and SSC consortia to the 
research programme reported here.

\bibliography{refs.bib}

\onecolumn

\begin{appendix}
\begin{center}
APPENDIX A : The Complete XGPS-I Source Catalogue.
\end{center}

%add J before source names! done
%source name positions should be truncated NOT rounded! done

%\small
%\footnotesize
\scriptsize

\begin{longtable}{lcccccccccccccl}

\vspace {0.05IN}

\\ \hline

XGPS-I &  &  \multicolumn{2}{c}{Equinox J2000} & \multicolumn{3}{c}{MOS count ks$^{-1}$} & & 
\multicolumn{3}{c} {pn count ks$^{-1}$} & & & \multicolumn{2}{c}{Optical} \\  

Designation & Obs & RA  & DEC  & s & h & b & q & s & h & b & q & HR & R & Identification \\ 

\hline

J182416-115554 & X1 & 18 24 16.63 & -11 55 54.5 & 0.7$\pm$0.7 & \bf 2.8$\pm$1.1 & \bf 3.4$\pm$1.3 & 0.51 & 0.0$\pm$2.1 & 3.7$\pm$2.2 & 1.7$\pm$2.6 & 1.00  & 0.72 & 19.4 & GPSR 19.236+0.495 \\ %135
J182426-115918 & X1 & 18 24 26.61 & -11 59 18.8 & \bf 3.9$\pm$1.1 & 0.0$\pm$0.7 & \bf 4.0$\pm$1.3 & 1.00 & \bf 18.7$\pm$3.7 & 0.9$\pm$1.7 & \bf 19.9$\pm$4.1 & 1.00  & -0.94 & 14.9 & U0750\_13092026 \\ %136
J182436-115127 & X1 & 18 24 36.20 & -11 51 27.5 & 0.3$\pm$0.4 & 0.5$\pm$0.5 & \bf 0.8$\pm$0.6 &  1.00 & 0.0$\pm$1.2 & 0.5$\pm$1.0 & 0.0$\pm$1.5 & 0.63  & 0.45 & - \\ %137
J182441-115903 & X1 & 18 24 41.19 & -11 59  3.5 & 1.1$\pm$0.6 & 1.4$\pm$0.6 & \bf 2.4$\pm$0.9 & 1.00 & 1.3$\pm$1.5 & 1.4$\pm$1.5 & 2.6$\pm$2.1 & 0.41  & 0.08 & - \\ %138
J182446-120429 & X1 & 18 24 46.45 & -12  4 29.5 & 0.0$\pm$0.7 & 0.0$\pm$0.6 & 0.0$\pm$0.9 & 0.02 & \bf 11.2$\pm$3.3 & 2.1$\pm$2.3 & \bf 13.5$\pm$4.0 & 0.98  & -0.69 & 9.8 & U0750\_13103886 \\
\end{longtable}

{\bf Key to Table}
\smallskip
\begin{description}
\item{Column 1} The X-ray source designation
\item{Column 2} The XGPS-I survey field in which the source was detected
\item{Column 3} The X-ray position
\item{Column 4} The derived MOS count rates in the soft (s), hard (h) and broad-bands
(b) with $\geq 5\sigma$ detections shown in bold and negative fluxes set to zero
\item{Column 5} The MOS {\it quality} flag
\item{Column 6} The derived pn count rates in the soft (s), hard (h) and broad-bands
(b) with $\geq 5\sigma$ detections shown in bold and negative fluxes set to zero
\item{Column 7} The pn {\it quality} flag
\item{Column 8} The X-ray hardness ratio HR (see text)
\item{Column 9} The R magnitude of the brightest optical source within a $6''$ error
circle based on the SuperCOSMOS digitisation of the R plate from the UK
Schmidt Survey
\item{Column 10} The designation of the optical source, if any, within a
$6''$ X-ray error circle. Here,  U = USNO, with all the other
references taken from the SIMBAD database.
\item
\item{Footnote:} The count rates quoted in this table are measured
count rates and have not been corrected for the signal loss due
to the limited size of the source cell (see \S 3.2).
\end{description}

\end{appendix}

\end{document}